\documentclass[prl,twocolumn,aps,showpacs,superscriptaddress]{revtex4-1}

\usepackage{graphicx}
\usepackage{url}

\begin{document}

\title{Heavy Weyl fermion state in CeRu$_4$Sn$_6$}

\author{Yuanfeng Xu}
\affiliation{Beijing National Laboratory for Condensed Matter Physics,
  and Institute of Physics, Chinese Academy of Sciences, Beijing
  100190, China}
\affiliation{Collaborative Innovation Center of Quantum Matter,
  Beijing, China}

\author{Changming Yue}
\affiliation{Beijing National Laboratory for Condensed Matter Physics,
  and Institute of Physics, Chinese Academy of Sciences, Beijing
  100190, China}
\affiliation{Collaborative Innovation Center of Quantum Matter,
  Beijing, China}

\author{Hongming Weng}
\affiliation{Beijing National Laboratory for Condensed Matter Physics,
  and Institute of Physics, Chinese Academy of Sciences, Beijing
  100190, China}
\affiliation{Collaborative Innovation Center of Quantum Matter,
  Beijing, China}

\author{Xi Dai}
\email{daix@iphy.ac.cn}
\affiliation{Beijing National Laboratory for Condensed Matter Physics,
  and Institute of Physics, Chinese Academy of Sciences, Beijing
  100190, China}
\affiliation{Collaborative Innovation Center of Quantum Matter,
  Beijing, China}

\date{\today}

\begin{abstract}
A new type of topological state in strongly corrected condensed matter systems, heavy Weyl fermion state, has been 
found in a heavy fermion material CeRu$_4$Sn$_6$, which has no inversion symmetry . Both two different types of Weyl points, type I and II, can be
found in the quasi-particle band structure obtained by  the LDA+Guztwiller calculations, which can treat the strong correlation
effects among the f-electrons from Cerium atoms. The surface calculations indicate that the topologically protected Fermi arc states 
exist  on the (010) but not on the (001) surfaces.
\end{abstract}
\maketitle
Recently different types of topological semimetals~\cite{TSM_review} have been proposed and observed in condensed matter systems, i.e. the Weyl semimetal(WSM) phase in transition metal compounds with the magnetic order, ~\cite{wan, HgCrSe} WSM phase in non-central symmetric crystals,~\cite{TaAs_Weng, HuangSM_Weyl, TaAs_arc, Xu2015a, TaAs_node, Yang2015} Dirac semimetal (DSM) phase in intermetallic compounds~\cite{Na3Bi, Cd3As2, Na3Biexp, Cd3As2exp, Cd3As2expCava, xu_observation_2014} and nodal line semimetal phase in anti-perovskite compounds.~\cite{Cu3NPd, Cu3NPdKane} In all the above-mentioned material systems, the electron-electron correlation effect is weak, and the band structure, as well as the existence of Weyl nodes, can be obtained quite accurately by density functional theory. On the other hand, the topological non-trivial electronic structure can be found in strongly correlated material systems as well, for instance, the topological Kondo insulator phase in SmB$_6$~\cite{Dzero:2010dj, Dzero:2012kx, SmB6_Lu2013} can be viewed as the strongly correlated $Z_2$ topological insulator, which has attracted lots of research interests in recent years.~\cite{Kim:2012dj, Jiang:2013tl, Xu:2013th, YbB6YbB12, Dzero:2016review} In SmB$_6$, the correlation effects generated by the strong Coulomb repulsive interaction among $f$-electrons suppress the bandwidth dramatically but leave the topological features of the electronic structure unchanged~\cite{Wang:2010tg, Wang:2012fj, SmB6_Lu2013, YbB6YbB12}.

In the present letter, we propose that CeRu$_4$Sn$_6$,~\cite{CeRu4Sn6_CryStr, CeRu4Sn6_kondo2010, CeRu4Sn6_optical2013, CeRu4Sn6_topological2015, CeRu4Sn6_dmft2016} a typical heavy fermion material, contains Weyl points in its quasiparticle band structure near the Fermi level and thus belongs to a new class of strongly correlated topological phase, heavy Weyl fermion state. Comparing to other WSMs found in non-interacting systems, the WSM phase in heavy fermion system has more fruitful physical properties due to the following reasons. Firstly, unlike the non-interacting systems, the heavy quasiparticle bands are fully developed only at the low temperature. Therefore, how the physics related to the topological electronic structure evolves as the decrement of temperature will become an crucial problem for the heavy Weyl fermion phase, which may lead to new unique phenomena in these systems. Secondly, in heavy fermion systems, the energy scale of the quasiparticle bands is orders smaller than the ordinary semiconductor or semimetal systems, which makes it more sensitive to various of the external field, i.e. the pressure, magnetic field and strain, providing large tunability to the distribution of the Weyl nodes. 

\begin{figure}[htp]
\includegraphics[clip,width=3.4in,angle=0]{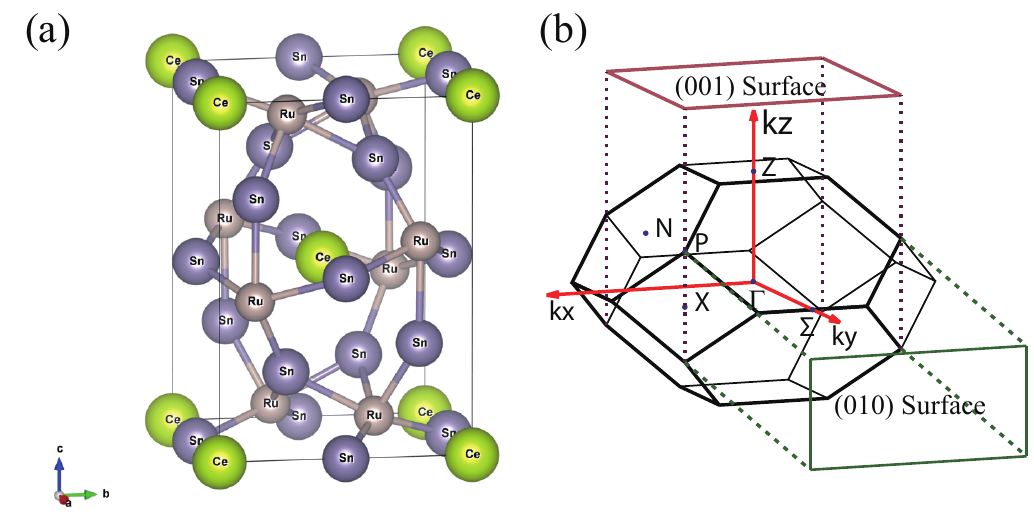}
\caption{(Color online) Crystal structure and Brillouin zone (BZ). (a) The crystal symmetry of CeRu$_4$Sn$_6$. (b) The bulk BZ and the projected surface BZ for both (001) and (010) surfaces.}\label{fig1}
\end{figure}

The crystal structure of  CeRu$_4$Sn$_6$~\cite{CeRu4Sn6_CryStr} has no inversion centre and the spin-orbit coupling (SOC) derived from the 4$f$ orbitals of Cerium and 4$d$ orbitals of Ru splits the quasi-particle bands, which makes it possible to have Weyl points near the Fermi level. The electronic structure of CeRu$_4$Sn$_6$ has been studied both theoretically and experimentally by several groups already.~\cite{ CeRu4Sn6_kondo2010, CeRu4Sn6_optical2013, CeRu4Sn6_topological2015, CeRu4Sn6_dmft2016} As introduced in references~\cite{CeRu4Sn6_optical2013, CeRu4Sn6_topological2015},  the band structure obtained by local density approximation (LDA) is semimetal type 
with the vanishing indirect but finite direct energy gap  between the valence band mostly consisting of Ruthenium 4$d$ orbitals and conduction band formed mostly by the Cerium 4$f$ orbitals. The LDA+DMFT calculation has been applied to this material by K. Held's group to capture the strong correlation effects.~\cite{CeRu4Sn6_dmft2016}  Besides the reduction of the bandwidth for the $f$-bands, another significant consequence of the correlation effect is to enhance the crystal splitting within the $J=5/2$ subspace and push down the lowest 4f bands with the $J=5/2, J_z=\pm1/2$ character, which leads to ``inverted features" between 4f and 4d bands in some area of the BZ mimicking the situation in SmB$_6$. This picture has been partly supported by the X-ray absorption and scattering data as discussed in reference~\cite{CeRu4Sn6_topological2015}. 

The previous studies imply that CeRu$_4$Sn$_6$ might be an another topological heavy fermion material. While unlike the situation of SmB$_6$, the renormalised quasiparticle band structure is not of the semiconductor but semimetal type. In the present letter, by LDA+Gutzwiller method,~\cite{LDA+GW_2009} we have studied the renormalised quasiparticle band structure of CeRu$_4$Sn$_6$ and confirmed that it contains Weyl points formed by heavy fermions. Further calculations on different surfaces indicate that due to the existence of the three dimensional bulk Fermi surfaces, protected surfaces states (SS) do not 
exist on the (001) surface, but they do exist on the (010) surface, where the projection of the bulk Fermi surfaces are well separated. Our LDA+Gutzwiller calculation then reveals very long and beautiful 
Fermi arc pattern on the (010) surface, which can be detected by various of experiments including the angle resolved photo emission (ARPES) and quantum oscillation.

CeRu$_4$Sn$_6$ is crystalized in body centered tetragonal lattice with space group of $I\bar{4}2m$ (No. 121), which has no inversion center(Fig. 1).
The experimental lattice constants $a$=6.8810~\AA, and $c$=9.7520~\AA~are adopted in our calculation~\cite{CeRu4Sn6_CryStr}. The Ce and Ru
 atoms are located at Wyckoff positions 2$a$ (0.0, 0.0, 0.0) and 8$i$ (0.82938, 0.82938, 0.42107), respectively. And the Sn atoms take 
 two Wyckoff positions 8$i$ (0.82134, 0.82134, 0.70476) and 4$c$ (0.0, 0.5, 0.0).
In the present letter, we first perform the electronic structure calculation by using Vienna $ab~initio$ Simulation Package (VASP)~\cite{VASP_1996} 
with the PBE-GGA type exchange correlation potential. The plane-wave cutoff energy is 410 eV.
Then the maximally localized Wannier functions for 4$f$ and 5$d$ orbitals on Ce, 4$d$ orbitals on Ru and  5$p$ 
orbitals on Sn atoms have been constructed by using the Wannier90 package.~\cite{wannier90}
 The SOC strength is obtained by fitting to the the corresponding full-relativistic first-principles calculation, which are
0.096 eV for Ce 4$f$, 0.160 eV for Ru 4$d$  and 0.202 eV for Sn 5$p$ orbitals.
The on-site interactions among the partially occupied $f$-orbitals are crucial, which can be properly treated by the LDA+Gutzwiller method~\cite{LDA+GW_2009, SmB6_Lu2013, YbB6YbB12}. We take the Coulomb interaction $U_d$ of 5.0 eV, and Hund's rule coupling $J_h$ of 0.818 eV, the full localized limit scheme is adopted for the double counting energy.

\begin{figure}[htp]
\includegraphics[clip,width=3.4in,angle=0]{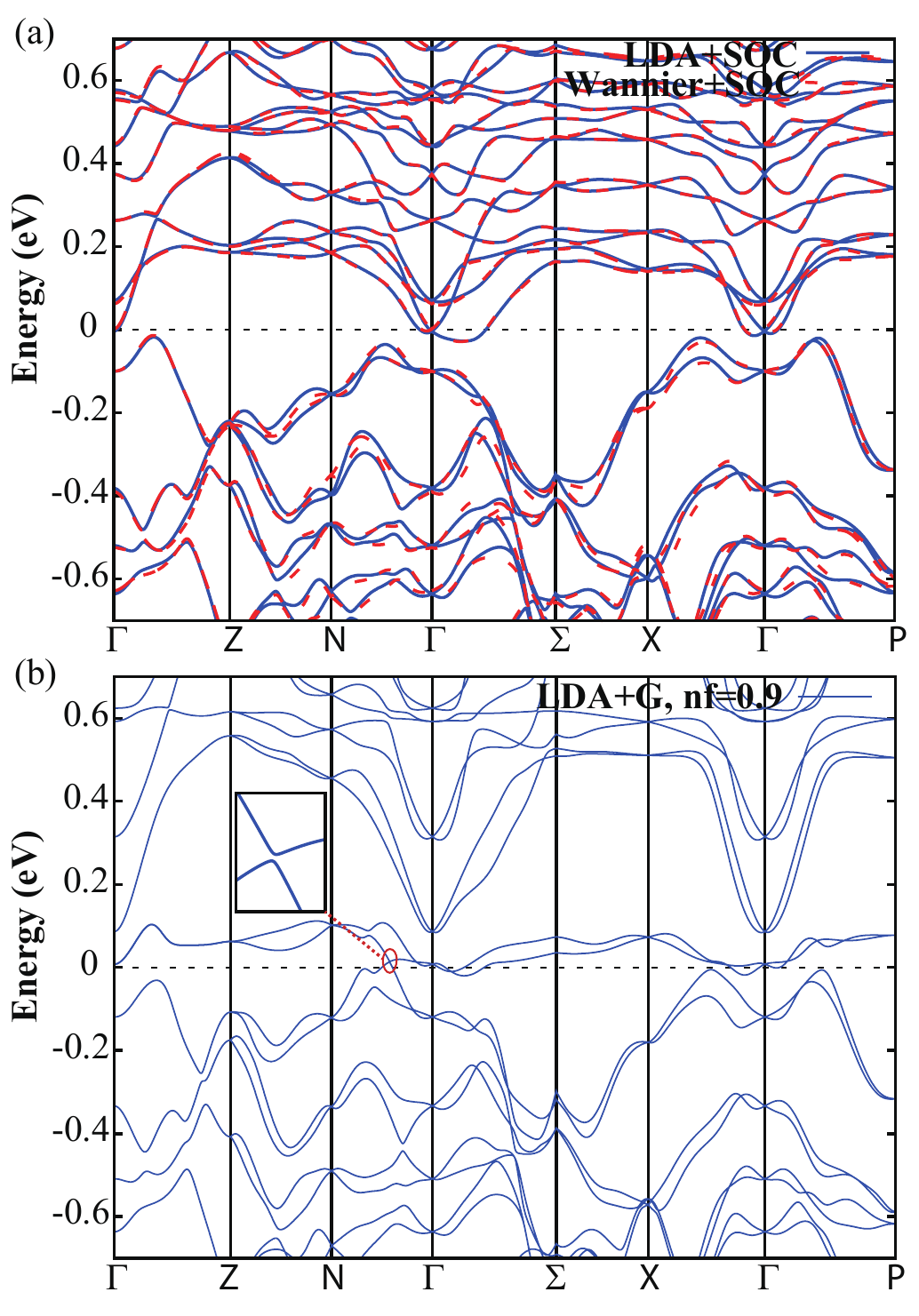}
\caption{(Color online) (a) The band structure of CeRu$_4$Sn$_6$ calculated by LDA+SOC (the solid line) and Wannier+SOC (the dash line). (b) The renormalized band structure of CeRu$_4$Sn$_6$ calculated by LDA+Gutzwiller method.The Fermi energy is set to 0. \label{Fig. 2}}
\end{figure}

The LDA band structure of CeRu$_4$Sn$_6$ is calculated and plotted in Fig. 2(a) with the SOC being fully considered,
which is semimetal type with very slight overlap between conduction and valence bands. The direct gap remains finite throughout the
whole BZ at the LDA level. Due to the strong repulsive
interaction among f-electrons, the quasiparticle bands mainly with 4f character will be strongly renormalised and only appear
below some certain temperature scale (the Kondo temperature), below which the f-electrons start to participate in the coherent motion.
In the present study, we apply the LDA+Gutzwiller method to quantitatively calculate the renormalised quasiparticle bands, which
are plotted in Fig. 2(b). Compared to the LDA band structure, there are two major corrections caused by the strong correlation effects.
Firstly, the total bandwidth of the 4f bands have been suppressed by approximately 2 times. Secondly and more importantly for this
particular material, the splitting among the 4f-orbitals has been greatly enhanced leading to two direct consequences, the bands with 
$|J=5/2;J_z=\pm1/2>$ character are pushed down to mix strongly with the 4d bands from the Ru atoms and at meanwhile the bands with 
$|J=7/2>$ character are pushed up to about 1.2 eV above the Fermi level. Compared with the previous DMFT study, our renormalised band 
structure is in very good agreement with their results. The histogram of the atomic configurations in the ground state can be obtained
by the LDA+Gutzwiller method and are plotted in Fig. 6(c), from which one can estimate the average occupation number of the
Cerium 4f orbitals to be 0.9. These results are quite consistent with the recent X-ray absorption data suggesting the maximum occupation
of the 4f orbitals to be 0.95. 

Due to the lack of inversion centre, all the quasi-particle bands are non-degenerate away from the eight time reversal invariance k-points.
Point group symmetry analysis indicates that the mixing of the $|J=5/2;J_z=\pm1/2>$ and the 4d states near the Fermi level will 
open hybridisation gap between them along any high symmetry lines as we can find so in Fig. 2(b). At generic k points, there is no
symmetry can protect the gapless nodes, but in three dimensional space, accidental degeneracy can appear between the energy bands without spin
degeneracy leading to the appearance of Weyl points (WP). WPs are monopoles with positive or negative
chirality for the Berry curvature calculated from all the occupied bands, which satisfies the Gauss's law. Therefore the integral of the Berry's 
curvature on any closed surface in the BZ will give us the total chirality of the WPs enclosed inside it, which can help us to confirm the
existence of the WP quickly. 

\begin{figure}[htp]
\includegraphics[clip,width=3.4in,angle=0]{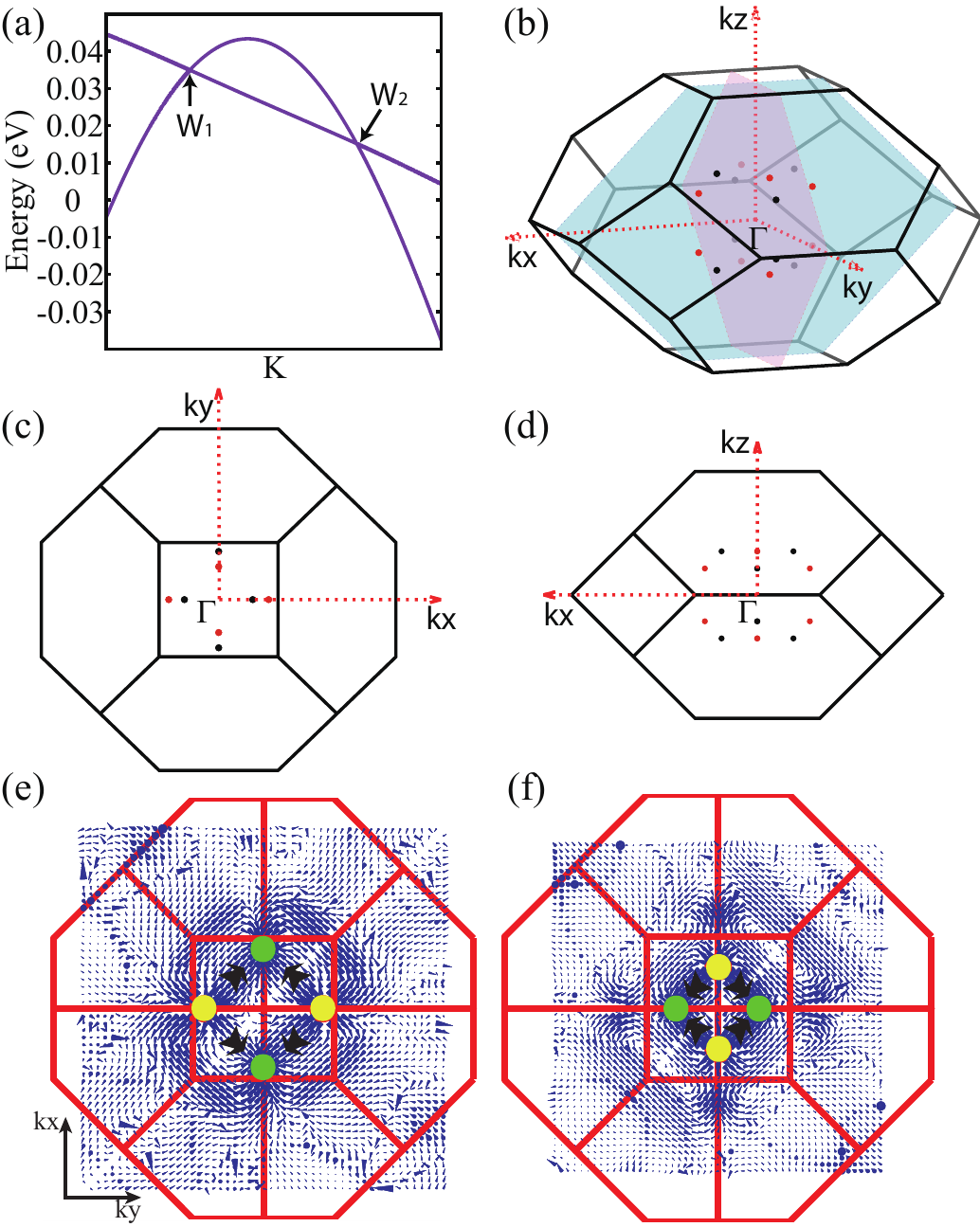}
\caption{(Color online) (a)The energy dispersion along the direction connecting $W_1$ and $W_2$.(b)3D view of the 8 pairs WPs in the BZ. (c)Top view from [001] and (d) side view from [010] directions for the WPs. (e)The distribution of Berry curvature for the $k_z$=0.139 \AA$^{-1}$ plane, where the green and yellow dots denote the Weyl points with negative and positive chirality. (f)same as (e) but for the $k_z$=0.227 \AA$^{-1}$ plane .}\label{fig3}
\end{figure}

\begin{table}
\caption{The position of the two nonequivalent Weyl points in the BZ. Fermi energy is set to 0.}
\begin{tabular}{c c c c}
\hline
\hline
Weyl point&Position(\AA$^{-1}$)&Chirality&$E$(meV)\\
\hline
$W_1$&~~(0.132, 0.0, 0.227)~~&-1&35\\
$W_2$&~~(0.193, 0.0, 0.139)~~&+1&15\\
\hline
\hline
\end{tabular}
\label{Tabel.I}
\end{table}
The precise positions for the WPs have been obtained by LDA+Gutzwiller calculation on a much dense grid in k-space. As shown in Fig. 3(e)(f), 
the WPs with positive and negative chirality can be identified as the ``source" and ``drain" of the Berry curvature.
Totally there are eight pairs of WPs as illustrated in Fig. 3(b)(c)(d) and Tabel. I, which can be divided into two groups labeled as $W_1$ and $W_2$ so that the 
WPs in each group can be linked by the crystal symmetry. The energy dispersion along the direction connecting $W_1$ and $W_2$ (as illustrated in
Fig. 3(b)) has been plotted in Fig. 3(a), which clearly indicates that $W_2$ is type II WPs whereas $W_1$ is type I. The coexistence of the two different
types of WPs has been discussed recently in reference \cite{Chengang:2016} and here we find the first realistic material which belongs to that category.
\begin{figure}[htp]
\includegraphics[clip,width=3.4in,angle=0]{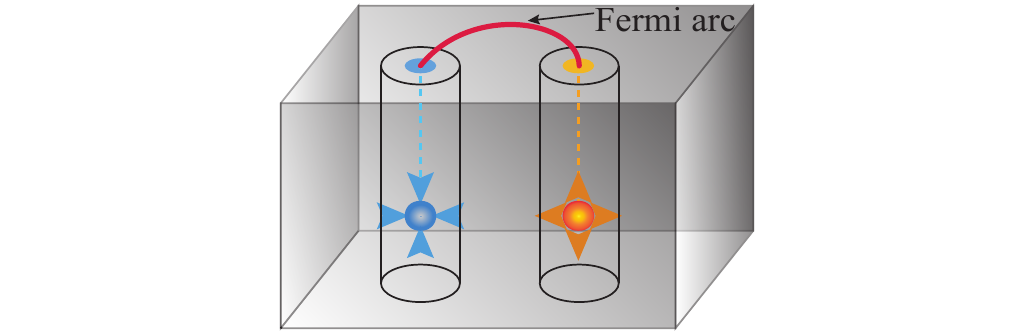}
\caption{(Color online) A pair of Weyl points with opposite chirality and the Fermi arc connect their projective points on the surface. }\label{fig4}
\end{figure}

In ideal WSM, the Fermi level only cuts the WPs and the FS of the bulk states are isolated points, which will generate protected Fermi arcs on any
surface as long as the projection points of the WPs on that particular surface BZ are separated from each other.~\cite{wan, HgCrSe} In realistic WSM materials, there are
finite size bulk FS  as well, which projects to the surfaces BZ as ellipses enclosing different projection points of the WPs. Then following the prove 
in reference~\cite{wan}, the topologically  protected Fermi arc still exists when these ellipses don't overlap on each other as schematically shown in Fig. 4.
In CeRu$_4$Sn$_6$, the finite bulk FSs do exist, which project to the surface BZ of (001) and (010) surfaces  as shown in Fig. 5. Interestingly although these projection area overlap on the (001) surfaces, they are well separated on the (010) surfaces, which leads to the protected Fermi arcs existing on the 
(010) but not the (001) surfaces.
\begin{figure}[htp]
\includegraphics[clip,width=3.4in,angle=0]{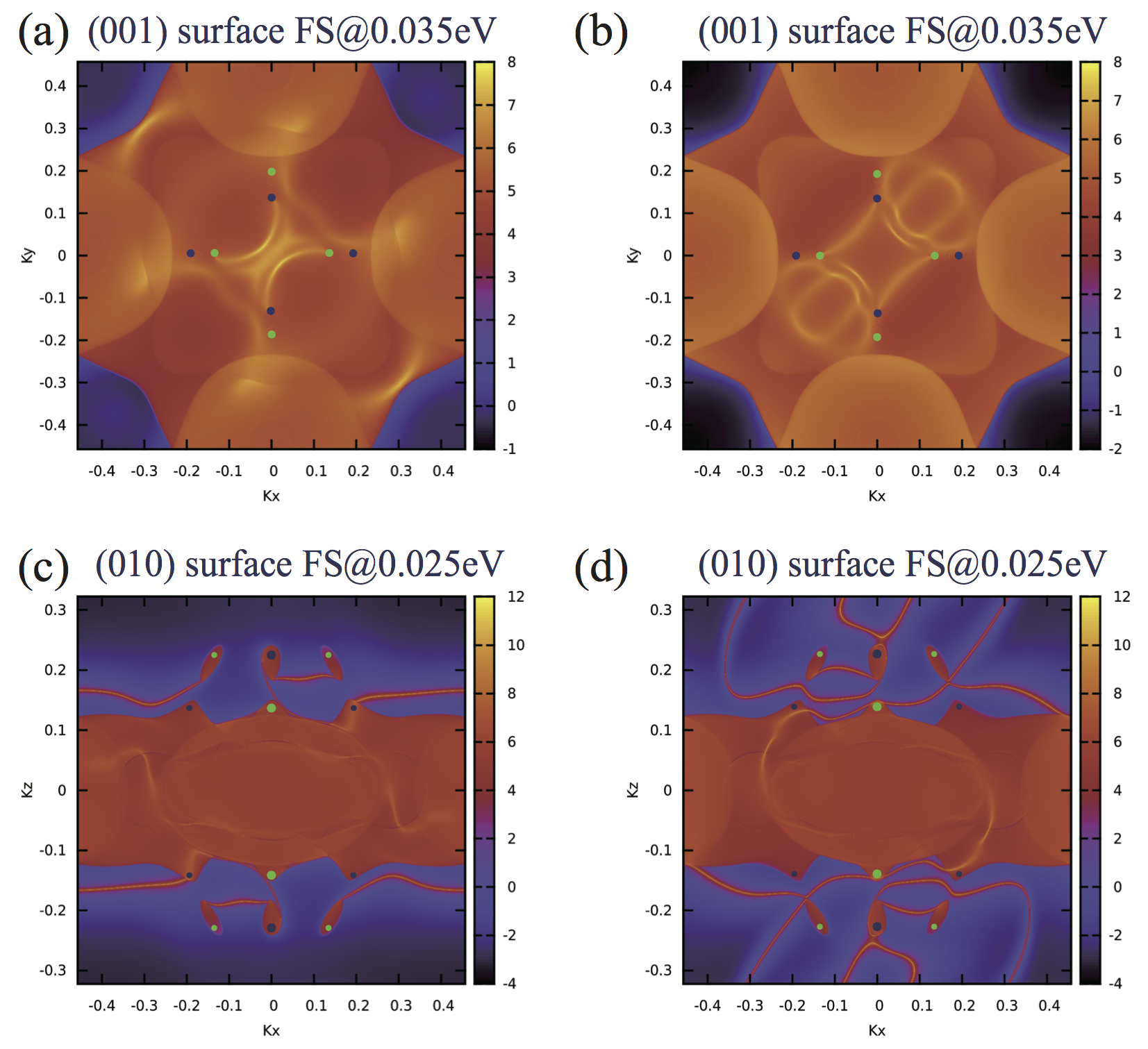}
\caption{(Color online) (a)(b) Fermi surface on (001) surface(both top and bottom surface of the sample). All the points represent two WPs with same Chirality projecting on the top of each other. (c)(d) Fermi surface on (010) surface. Small dot represents the single projected WPs, and large dot represents two WPs with same Chirality projecting on the top of each other.  The coordinates are in units of \AA$^{-1}$. }\label{fig5}
\end{figure}
We further calculated the SS on both (001) and (010) surfaces with the software package Wannier\_tools \cite{Wu_wanntools} using the renormalised effective tight binding Hamiltonian obtained by LDA+Gutzwiller together with
the Green's function method. As shown in Fig. 5(a)(b), there are no clear SS on the (001) surface due to the overlap of the bulk FS. While for the surfaces along
the (010) direction, the situation is very different, there are very long and clear Fermi arcs connecting WPs with opposite chirality as plotted in Fig. 5(c)(d).
These unique SS on (010) surfaces will lead to abnormal physical properties, which are yet to be detected by ARPES, transport and quantum oscillation experiments.

The total number and location of the WPs  are quite sensitive to the detail of the band structures. In the present studies, although the interaction parameters 
as well as the double counting scheme adopted in the LDA+Gutzwiller calculations are all reasonable leading to consistent results with the previous numerical and
experimental studies, small change of the band structure may still lead to qualitative change of the WPs. Therefore it is important to further check the 
robustness of the heavy Weyl fermion state against the small uncertainty from the numerical calculations, which is unavoidable at the current stage. Compared to
the change of interaction parameters, the quasi-particle band structure is way more sensitive to the change of double counting potential, which slightly modifies the 
physical valence of the Cerium ions. Therefore in the present paper, we calculated different quasi-particle band structure with the modification of the double
counting potential and plotted the phase diagram with the resulting occupation of the 4f orbitals in Fig. 6(d). Our results indicate that the existence of the heavy
Weyl fermion state in this material is very robust, it appears when $n_f>0.87$. For $n_f>0.92$, the number of WPs increases to be 12 pairs leading to more
complicated fermi arc patterns on the surfaces. 

\begin{figure}[htp]
\includegraphics[clip,width=3.4in,angle=0]{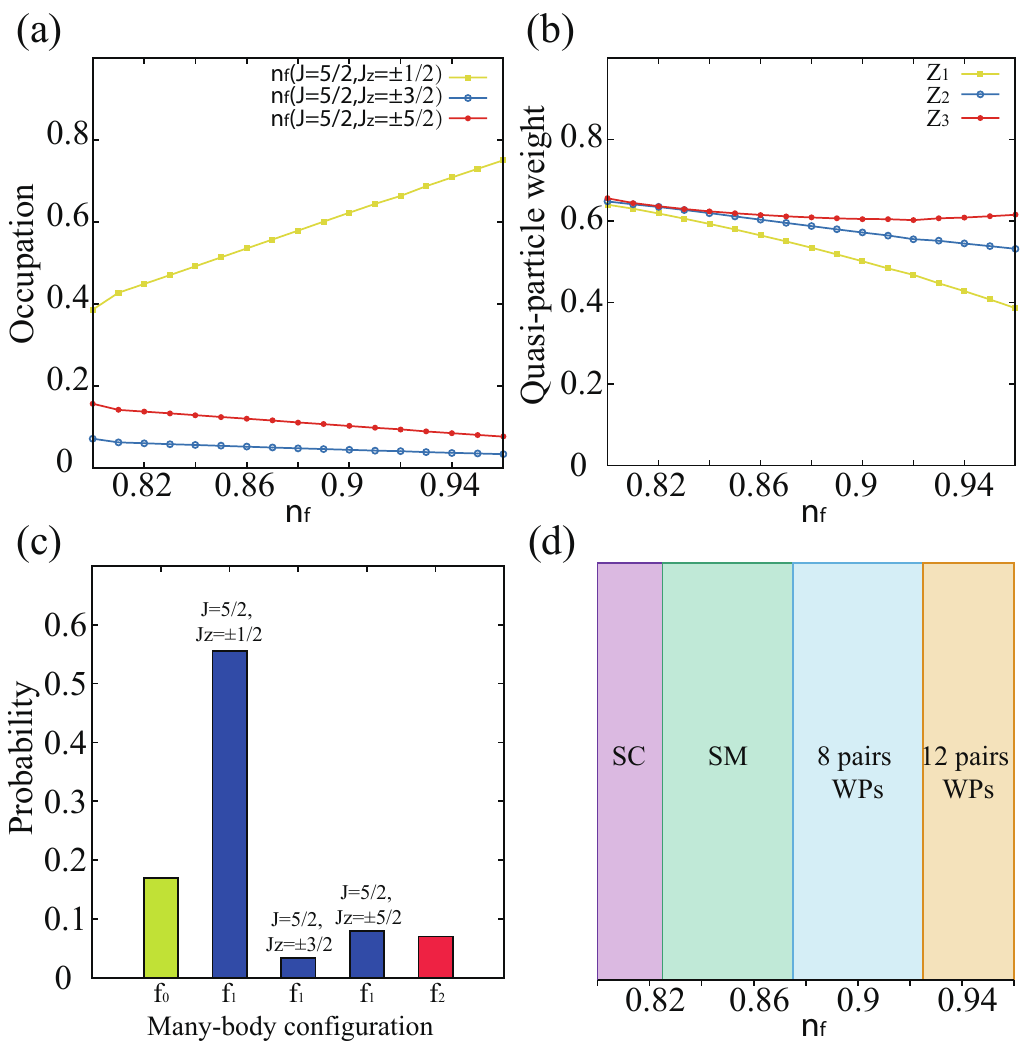}
\caption{(Color online) (a)The occupation of f orbitals with $J=5/2$, $J_z=\pm1/2,\pm3/2,\pm5/2$ character; (b)the quasi-particle weights for 
the 4f orbitals with $J=5/2$; (d)phase diagram of CeRu$_4$Sn$_6$ with the change of the total 4f orbital occupation number. (c)The histogram of the atomic configurations for the case of $n_f=0.9$.}\label{fig6}
\end{figure}

The above phase diagram also indicates that comparing to the WSM state found in weakly correlated materials, the properties of the heavy Weyl fermion state introduced here
is much sensitive to the external fields that can modify the effective valence of Cerium, for instance the pressure, strain and chemical doping, which provides
great tunability in these systems and make it a promising material platform for follow up studies on the relationship between correlation and topology.

In conclusion, based on the LDA+Gutzwiller calculation, we find that CeRu$_4$Sn$_6$ is the first Weyl semimetal in heavy fermion materials. 
The unique electronic structure of CeRu$_4$Sn$_6$ is greatly renormalized by the strong correlation effects among the f-electrons, leading to
the appearance of WPs in the heavy quasiparticle bands. The surface calculations indicate that the Fermi arcs on the (010) surface are well
separated from the projection of the bulk bands and can be thus detected by ARPES or quantum oscillation experiments.

We acknowledge the supports from National Natural Science Foundation of China (Grant Nos. 11274359 and 11422428),
the National 973 program of China (Grant No. 2013CB921700), the ``Strategic Priority Research Program (B)"
of the Chinese Academy of Sciences (Grant No. XDB07020100) and MOST project under the contract number 2016YFA0300604. Partial of the calculations were preformed on TianHe-1(A), the National Supercomputer Center in Tianjin, China.
\bibliography{CRS_ref}

\end{document}